\documentstyle[12pt]{article}
\title{Formal BFV-Type Representation of Path-Integral for Dynamical
System with Second Class Constraints}
\author{A.A. Deriglazov\thanks{E-mail: deriglaz@phys.tsu.tomsk.su}}
\date{Department of Mathematical Physics,\\
Tomsk Polytechnical University,\\
634004 Tomsk, Russia}
\def\ve{\varepsilon}
\def\det{\mathop{\rm det}\nolimits}
\def\CP{{\cal P}}
\def\gh{\mathop{\rm gh}\nolimits}
\def\pa{\partial}
\begin{document}
\large

\maketitle
\begin{abstract}
It is shown that the phase space of a dynamical system subject to
second class constraints can be extended by ghost variables in such a
way that some formal analogies of the $\Omega$-charge and the unitarizing
Hamiltonian can be constructed. Then BFV-type path integral
representation for the generating functional of Green's functions is
written and shown to coincide with the standard one.
\end{abstract}

It is known that an embedding of second class constraints into the
Hamiltonian BFV-quantization scheme can be done in several different
ways: by application of the conversion methods \cite{1, 2}, by passing
to holomorphic representation of the constraints \cite{3, 4}, and
within the framework of the unified constrained dynamics \cite{5}.
Unfortunately, some restrictions on a structure of the initial
constraint system must be imposed in each of these schemes. By this
reason, a direct application of the developed methods in a manifestly
Poincar\'e covariant fashion turns out to be problematic for some
concrete models (see \cite{6, 7} and references therein). In
particular, for those cases the problem of constructing a manifestly
covariant and really calculable expression for the generating
functional of Green's functions seems to have no a fully satisfactory
solution in the path integral quantization context [8--10].
The purpose of this letter is to demonstrate that for a dynamical
system subject to second class constraints $G_\alpha\approx0$, some
formal quantities analogous to the $\Omega$-charge and the unitarizing
Hamiltonian can be constructed. Then the standard expression for the
path integral \cite{11}
\begin{equation}
Z=\int dz\,\det^{1/2}\{G_\alpha,G_\beta\}\delta(G)\exp i\int d\tau
(p_A\dot q^A-H_0)
\end{equation}
can be rewritten in BFV-like form (see Eqs. (21), (22) below). It allows
to hope that this formal construction will be useful in attempts to
treat quantization of second class constraints on the same footing as
that of first class ones.

Consider a dynamical system with phase space variables $z^A\equiv(q^A,
p_A)$, Hamiltonian $H_0(z)$ and second class constraints
$G_\alpha(z)\approx0$:
\begin{eqnarray}
&& S_H=\int d\tau(p_A\dot q^A-H_0+\lambda^\alpha G_\alpha),\cr
&& \{G_\alpha,G_\beta\}_{PB}=\Delta_{\alpha\beta}(z), \qquad
\det\Delta\ne0,\\
&& \{G_\alpha,H_0\}_{PB}={V_\alpha}^\beta(z)G_\beta.\nonumber
\end{eqnarray}
For definiteness, all the phase space coordinates and constraints are
supposed to be even: $\ve(z^A)=\ve(G_\alpha)=0$.
\paragraph{\large\bf A constructing of the $\Omega$-charge.} To get the
quantity similar to the $\Omega$-charge, with property
$\{\Omega,\Omega\}_{PB}=0$, let me extend the initial phase space in
the following way:
\begin{equation}
(q^A,p_A), \,\lambda^\alpha, (C^\alpha,{\cal P}_\alpha),\, D^\alpha,\,
(\alpha,\beta),\,(\eta,\nu)
\end{equation}
where the brackets contain canonically conjugate variables, while
$\lambda^\alpha$ and $D^\alpha$ are assumed to be dynamically passive.
The Grassmann parity, ghost numbers and nonvanishing Poisson brackets
of the variables are
\begin{equation}
\begin{array}{l}
\ve(C^\alpha)=\ve(\CP_\alpha)=\ve(\alpha)=\ve(\beta)=\ve(D^\alpha)=1,\\
\ve(\eta)=\ve(\nu)=0;\\
\gh C^\alpha=-\gh\CP_\alpha=\gh\beta=-\gh\alpha=\gh\eta=-\gh\nu=\gh
D^\alpha=+1;\\
\{q^A,p_B\}={\delta^A}_B;\quad \{C^\alpha,\CP_\beta\}=-{\delta^\alpha}_\beta,
\quad \{\alpha,\beta\}=-1, \quad \{\eta,\nu\}=1.\end{array}
\end{equation}
Note that in contrast to conversion methods, no additional variables
for conversion of second class constraints into effective first class ones
has been introduced. Instead of this, the $\Omega$-charge will be constructed
directly in terms of the initial constraints.

The following notation for the expansion terms of a phase space
function $X(C^\alpha,\dots)=\sum_{n=0}^\infty X_n$ in powers of the
ghost parameters $C^\alpha$
\begin{equation}\begin{array}{l}
X_n\equiv C^{\alpha_1}\dots C^{\alpha_n}X_{n,\alpha_1\dots\alpha_n},\\
X_{[n]}\equiv\displaystyle\sum_{k=0}^n C^{\alpha_1}\dots C^{\alpha_k}
X_{k,\alpha_1\dots\alpha_k},\end{array}
\end{equation}
will be used, and analogously for function $X(\alpha,C^\alpha,\dots)\equiv
\sum_{n=0}^\infty(X_{n,0}+\alpha X_{n,1})$.

Introducing now the function $\Omega_{[1]} = C^\alpha G_\alpha$ one
has: $\{\Omega_{[1]},\Omega_{[1]}\}=$\linebreak $=C^\alpha C^\beta\Delta_{\alpha\beta}
(z)\equiv W$. Then, by virtue of the Jacobi identity, the quantity
\begin{equation}
\Omega=\Omega_{[1]}+\frac 1{\sqrt{2}}\alpha W + \frac 1{\sqrt{2}}\beta
\end{equation}
has the needed property $\{\Omega,\Omega\}=0$ (compare Eq. (6) with
that of Refs. \cite{3, 4}, where for the case of constraint system in
the holomorphic representation a similar construction for the
$\Omega$-charge was considered and its cogomologies were investigated).

\paragraph{\large\bf A constructing of the unitarizing Hamiltonian.} The
Hamiltonian $H$ is sought as a solution of the equation
\begin{equation}
\{\Omega,H\}=0,
\end{equation}
subject to conditions
\begin{equation}
H|_{\alpha=\beta=C=\CP=0}=H_0,\qquad \ve(H)=0,\qquad \gh H=0.
\end{equation}
The solution will be sought in a form of series expansion in powers of
the ghost parameters
\begin{eqnarray}
&& H=\sum_{n=0}^\infty(H_{n,0}+\alpha H_{n,1}),\cr
&& H_{n,0}=C^{\alpha_1}\dots C^{\alpha_n}(H_{n,0\alpha_1\dots\alpha_n}
{}^{\beta_1\dots\beta_n}\CP_{\beta_1}\dots\CP_{\beta_n}+\cr
&&\qquad +\beta\tilde H_{n,0\alpha_1\dots\alpha_n}{}^{\beta_1\dots
\beta_{n+1}}\CP_{\beta_1}\dots\CP_{\beta_{n+1}}),\\
&& H_{n,1}=C^{\alpha_1}\dots C^{\alpha_n}(H_{n,1\alpha_1\dots\alpha_n}
{}^{\beta_1\dots\beta_{n-1}}\CP_{\beta_1}\dots\CP_{\beta_{n-1}}+\cr
&&\qquad +\beta\tilde H_{n,1\alpha_1\dots\alpha_n}{}^{\beta_1\dots
\beta_n}\CP_{\beta_1}\dots\CP_{\beta_n}),\nonumber
\end{eqnarray}
where the structure functions $H_{n,i}$ can be determined step by step
after the substitution of Eq. (9) into Eq. (7).

Let me prove the existence theorem for the solution of Eqs. (7) and
(8), inductively in powers of $(C)^n$. It is not difficult to find a
solution in the lowest orders $n=1$ and $n=2$:
\begin{equation}
H_{[2]}=H_0-C^\alpha{V_\alpha}^\beta\CP_\beta-\sqrt{2}\alpha C^\alpha
C^\beta\{G_\alpha,{V_\beta}^\gamma\}\CP_\gamma,
\end{equation}
then
\begin{equation}
\{\Omega,H_{[2]}\}_{[2]}=0,\quad \mbox{or equivalently} \quad
\{\Omega,H_{[2]}\}=X_3+\dots\,.
\end{equation}

Now, suppose that a solution of the problem is known in $(C)^n$ order,
so that $H_{[n]}$ is obtained, and
\begin{equation}
\{\Omega,H_{[n]}\}=X_{n+1}+\dots\,.
\end{equation}
In order to find $H_{n+1}$, note that $(C)^{n+1}$ order in Eq. (7) has
the following structure:
\begin{equation}
\begin{array}{c} \displaystyle\{\Omega,H\}_{n+1}=-\Big(\delta+
\frac 1{\sqrt{2}}\frac \pa{\pa\alpha}\Big)H_{n+1}+X_{n+1}=0,\\
X_{n+1}\equiv\{\Omega,H_{[n]}\}_{n+1},\end{array}
\end{equation}
where $\delta\equiv G_\alpha\frac\pa{\pa\CP_\alpha}$ is an analog of
the Koszul--Tate differential for second class constraints. Inserting
the expansions $H_{n+1}=H_{n+1,0}+\alpha H_{n+1,1}$,
$X_{n+1}=X_{n+1,0}+\alpha X_{n+1,1}$ into Eq. (13), one gets that the
following two equations must be fulfilled separately
\begin{eqnarray}
&& \delta H_{n+1,1}=-X_{n+1,1},\\
&& \delta H_{n+1,0}=X_{n+1,0}-\displaystyle\frac 1{\sqrt{2}} H_{n+1,1}.
\end{eqnarray}
To investigate these equations one can use the well known properties of
the $\delta$-operator \cite{12, 13}: Let the initial constraints are of
the form $G_\alpha=p_\alpha-f_\alpha(q^A,p_i)$, where $p_\alpha$ are
part of momenta $p_A$ (which always can be done). Then any regular
solution of the equation $\delta X=0$ obeying $X|_{\CP=0}=0$ is that
$X=\delta K$. Further, a necessary and sufficient condition for the
existence of solutions to the inhomogeneous equation $\delta X = Y$
with the unknown $X$ is that $\delta Y =0$.

To be convinced that $\delta~{\rm[r.h.s.~of~Eqs.~(14),~(15)]}=0$,
note that the $(n+1)$-th order of the Jacobi identity
$\{\Omega,\{\Omega,H_{[n]}\}\}=0$ leads to the equation
\begin{eqnarray}
&\{\Omega,\{\Omega,H_{[n]}\}\}_{n+1}=G_\alpha\{C^\alpha,X_{n+1}\}+
\displaystyle\frac1{\sqrt{2}}\{\beta,X_{n+1}\}=\cr
&=-\displaystyle\left(\delta+\frac 1{\sqrt{2}}\frac\pa{\pa\alpha}\right)
X_{n+1}=-\delta X_{n+1,0}+\alpha\delta X_{n+1,1}-\frac 1{\sqrt{2}}
X_{n+1,1}=0,
\end{eqnarray}
or equivalently
\begin{eqnarray}
&& \delta X_{n+1,1}=0,\\
&& \delta X_{n+1,0}=-\displaystyle\frac 1{\sqrt{2}}X_{n+1,1}.
\end{eqnarray}
From Eq. (17) it follows that there exists a solution $H_{n+1,1}$ of
Eq. (14). Assuming it is found and substituting it into Eq. (15), one
gets, by virtue of Eqs. (14), (18): $\delta$[r.h.s. of Eq. (15)] = 0,
which proves the existence of a solution of Eq. (15) too. In the
result, the solution of the problem (7), (8) has been found in the
$(C)^{n+1}$ order.

To conclude the subsection, let me discuss an ambiguity in determining
the Hamiltonian $H$, which has two sources:

(i) By construction $H_{n+1}$ is found from equation of the form
$\delta X = Y$, solution of which is not unique: given a solution $X$
the quantity $X+\delta K$ is a solution also.

(ii) In the initial formulation, one could start from equivalent
constraints $G'_\alpha$ instead of $G_\alpha$:
$G_\alpha={d_\alpha}^\beta G'_\beta$, $\det d \ne0$ (note that I have
used the properties of the $\delta$-operator, which were formulated for
the constraint system of a special structure only).

It can be shown that the operator $\tilde\delta\equiv\left(\delta+
\frac 1{\sqrt{2}}\frac\pa{\pa\alpha}\right)$, which first appeared in Eq.
(13), has the same properties: $\tilde\delta X=0\Rightarrow
X=\tilde\delta K$, $K=K_0+\alpha K_1$, as the $\delta$-operator has.
Using this fact, one can repeat the standard reasoning [13] for
describing the ambiguity in determining $H$, and the final answer looks
as follows: let $H=H_0+\dots$ and $\Omega$ be a solution of Eq. (7),
where the $\Omega$ is given in the form (6). Then any another solution
$H'=H_0+\dots$ and $\Omega'$ (in particular $\Omega'$ may be
constructed with the help of equivalent constraints: $\Omega'=C^\alpha
G'_\alpha+\dots$) is related to the initial one as follows:
\begin{equation}
\begin{array}{l} H'=e^{\hat X}(H+\{\Omega,Y\}), \qquad \Omega'=e^{\hat
X}\Omega;\\
\hat XA\equiv\{X,A\},\end{array}
\end{equation}
with some $X$ and $Y$ obeying $\ve(X)=0$, $\gh\,X=0$, $\ve(Y)=1$, $\gh
Y=-1$.

\paragraph{\large\bf BFV-type representation for the path integral.}
Since within the framework of the presented construction part of
variables (namely $\lambda^\alpha$ and $D^\alpha$) are dynamically
passive, the following trivial extension of the BFV-theorem \cite{14, 15}
will be suitable for building the generating functional of Green's
functions: Let the extended phase space variables are splitted onto
two groups: $Z=(Q^A,P_A; \omega^a)$ where $\omega^a$ are dynamically
passive. Let $\{\Omega,\Omega\}=0$, $\{\Omega,H\}=0$, where the
$\Omega$-charge is independent on $\omega^a$. Then the expression
\begin{equation}
Z_\Psi=\int dZ \exp i\int_{\tau_1}^{\tau_2}d\tau [P_A\dot Q^A-H+
\{\Psi,\Omega\}]
\end{equation}
is independent on change of $\Psi$ provided the standard choice of the
BRST-invariant boundary conditions \cite{12} on $Q^A, P_A$-variables
has been done.

This proposition can be proved along the same lines as the standard one.

For the case under consideration one has
\begin{eqnarray}
&& Z_\Psi=\displaystyle\int dZ \exp i\int d\tau[p_A\dot q^A+\CP_\alpha
\dot C^\alpha +\beta\dot\alpha +\eta\dot\nu-H+\{\Psi,\Omega\}];\\
&& \Omega=C^\alpha G_\alpha +\displaystyle\frac \alpha{\sqrt{2}}
C^\alpha C^\beta\Delta_{\alpha\beta}+\frac 1{\sqrt{2}}\beta,\cr
&& H=H_0-C^\alpha{V_\alpha}^\beta\CP_\beta-\sqrt{2}\alpha C^\alpha
C^\beta\{G_\alpha,{V_\beta}^\gamma\}\CP_\gamma +\\
&& \qquad +O(C^3\CP^3, C^3\CP^4\beta, \alpha C^3\CP^2, \alpha\beta C^3
\CP^3).\nonumber
\end{eqnarray}

Now, let me demonstrate that $Z_\Psi$ in Eq. (21) exactly coincides
with the standard expression of generating functional for a dynamical
system with second class constraints that has been written in Eq. (1).
The ``gauge fixing fermion'' of the form
\begin{equation}
\Psi = -\frac 1\ve\lambda^\alpha\CP_\alpha-\sqrt{2}\beta-\frac 1\ve
\alpha\CP_\alpha D^\alpha,
\end{equation}
where $\ve$ is some numerical parameter, turns out to be suitable for
this aim. Substituting Eqs. (22), (23) into Eq. (21) and making use of
the following displacement of integration variables
\begin{equation}
\begin{array}{cc}
\alpha\to\ve\alpha, & \beta\to\displaystyle\frac 1\ve\beta;\\
\CP_\alpha\to\ve\CP_\alpha, &
\lambda^\alpha\to\ve\lambda^\alpha\end{array}
\end{equation}
with the unit Jacobian, one gets
\begin{eqnarray}
Z_\Psi&=&\displaystyle\int dZ \exp i\int d\tau\Big[p_A\dot q^A+\CP_\alpha
\Big(\frac1{\sqrt{2}}D^\alpha+\ve\dot C^\alpha-\ve C^\beta
{V_\beta}^\alpha\Big)+\cr
&+&\beta\dot\alpha+\eta\dot\nu-H_0+O(\ve^2\alpha C^2\CP^1)+
(\lambda^\alpha-\alpha D^\alpha)G_\alpha+\cr
&+&C^\alpha C^\beta\Delta_{\alpha\beta}-\sqrt{2}\ve\alpha\lambda^\alpha
C^\beta\Delta_{\alpha\beta}\Big].
\end{eqnarray}
Since $Z_\Psi$ is independent on a change of $\Psi$ and, as a
consequence, on a change of $\ve$, let me pass to the limit $\ve\to0$.
After that, by making use of the displacement $\lambda^\alpha\to
\lambda^\alpha+\alpha D^\alpha$ with the unit Jacobian, one obtains the
expression for $Z_\Psi$, in which the integrations over $\CP_\alpha$,
$D^\alpha$ can be immediately performed. The integrations over $\alpha,
\beta,\eta,\nu$ variables are performed by transition on discrete
lattice, and after the necessary regularization \cite{16}, one obtains
\begin{equation}
\int d\alpha d\beta d\eta d\nu \exp i\int d\tau(\beta\dot\alpha+
\eta\dot\nu)=\lim_{\varphi\to0}\int d\alpha(0)d\nu(0)\exp i(\varphi\alpha
-\varphi^2\pi\nu^2)=1.
\end{equation}
(Thus an introduction of the ghosts $\eta,\nu$ technically yields the
correct balance between the divergent integral contributions.) The
resulting expression for $Z_\Psi$ is simply coincides with Eq. (1), as
has been stated.

Note that the generating functional $Z_\Psi$ is independent on the
natural arbitrariness in constructing of $H$ and $\Omega$ being
described by Eq. (19). It can be proved following the same arguments as
in Ref. 13.

To conclude, in this article for a dynamical system subject to second
class constraints some specific extension of the initial phase space by
ghost variables has been suggested. It allows to construct formal
BFV-type representation (21), (22) for the path integral in a similar
framework as for the case of first class constraints. The only unusual
property of the construction is that the ghost number of the ``gauge
fixing fermion'' (23) is not fixed.

The author is sincerely grateful to P.M. Lavrov for helpful discussions.

\end{document}